\documentclass[aps,prb,twocolumn,showpacs,superscriptaddress,longbibliography]{revtex4-2}

\usepackage{amssymb}
\usepackage{amsfonts}
\usepackage{amsmath}
\usepackage{subfigure}
\usepackage{multirow}
\usepackage{bm}
\usepackage{textcomp}
\usepackage{color}

\usepackage{array}
\usepackage[table,xcdraw]{xcolor}
\usepackage{graphicx}
\usepackage{float}
\usepackage{parskip}
\usepackage{svg}
\usepackage[T1]{fontenc}
\usepackage[a4paper=true,pagebackref=false]{hyperref}
\hypersetup{colorlinks, citecolor=blue, filecolor=blue, linkcolor=blue, urlcolor=blue}

\usepackage[normalem]{ulem}
\usepackage[english]{babel}
\usepackage{url}
\usepackage{appendix}

\usepackage{tikz,xcolor,hyperref}

\definecolor{lime}{HTML}{A6CE39}
\DeclareRobustCommand{\orcidicon}{%
	\begin{tikzpicture}
		\draw[lime, fill=lime] (0,0) circle [radius=0.16]
		node[white] {{\fontfamily{qag}\selectfont \tiny ID}};
		\draw[white, fill=white] (-0.0625,0.095) circle [radius=0.007];
	\end{tikzpicture}%
	\hspace{-2mm}
}

\foreach \x in {A, ..., Z}{%
	\expandafter\xdef\csname orcid\x\endcsname{\noexpand\href{https://orcid.org/\csname orcidauthor\x\endcsname}{\noexpand\orcidicon}}
}

\begin{document}
	
	\title{Ferromagnetic Resonance in a Magnetically Dilute Percolating Ferromagnet: An Experimental and Theoretical Study}
	
	\author{\mbox{ Y.K. Edathumkandy\orcidA}}
	\affiliation{\href{https://ror.org/000sfad56}{Institute of Physics}, Polish Academy of Sciences, Aleja Lotników 32/46, PL-02668 Warsaw, Poland} 

	\author{\mbox{K. Das\orcidB}}
	\affiliation{\href{https://ror.org/000sfad56}{Institute of Physics}, Polish Academy of Sciences, Aleja Lotników 32/46, PL-02668 Warsaw, Poland}
	
	\author{\mbox{K. Gas\orcidC}}
	\affiliation{\href{https://ror.org/000sfad56}{Institute of Physics}, Polish Academy of Sciences, Aleja Lotników 32/46, PL-02668 Warsaw, Poland}
	\affiliation{Center for Science and Innovation in Spintronics, \href{https://ror.org/01dq60k83}{Tohoku University}, 2-1-1 Katahira, Aoba-ku, Sendai 980-8577, Japan}

	\author{\mbox{D. Sztenkiel\orcidD}}
	\affiliation{\href{https://ror.org/000sfad56}{Institute of Physics}, Polish Academy of Sciences, Aleja Lotników 32/46, PL-02668 Warsaw, Poland}
	
	\author{\mbox{D. Hommel\orcidE}}
	
	\affiliation{\href{https://ror.org/03rvn3n08}{Łukasiewicz Research Network - PORT Polish Center for Technology Development}, Stabłowicka 147, Wrocław 54-066, Poland}
	\affiliation{\href{https://ror.org/04xdyc983}{Institute of Low Temperature and Structure Research}, Polish Academy of Sciences, Okólna 2, Wrocław 50-422, Poland}

	\author{\mbox{H. Przybylińska\orcidF}}
	\affiliation{\href{https://ror.org/000sfad56}{Institute of Physics}, Polish Academy of Sciences, Aleja Lotników 32/46, PL-02668 Warsaw, Poland}
	
	\author{\mbox{M. Sawicki\orcidG}}
	\affiliation{\href{https://ror.org/000sfad56}{Institute of Physics}, Polish Academy of Sciences, Aleja Lotników 32/46, PL-02668 Warsaw, Poland}
	\affiliation{Laboratory for Nanoelectronics and Spintronics, Research Institute of Electrical Communication,\href{https://ror.org/01dq60k83}{Tohoku University}, 2-1-1 Katahira, Aoba-ku, Sendai 980-8577, Japan}
	
	\begin{abstract}
		
Ferromagnetic resonance (FMR) serves as a powerful probe of magnetization dynamics and anisotropy in percolating ferromagnets, where short-range interactions govern long-range magnetic order. 
We apply this approach to Ga$_{1-x}$Mn$_x$N ($x \simeq 8$\%), a dilute ferromagnetic semiconductor, combining FMR and superconducting quantum interference device magnetometry. 
Our results confirm the percolative nature of ferromagnetism in (Ga,Mn)N, with a Curie temperature $T_{\mathrm{C}} = 12$~K, and reveal that despite magnetic dilution, key features of conventional ferromagnets are retained.
FMR measurements establish a robust uniaxial anisotropy, dictated by Mn$^{3+}$ single-ion anisotropy, with an easy-plane character at low Mn content. 
While excessive line broadening suppresses FMR signals below 9 K, they persist up to 70~K, indicating the presence of non-percolating ferromagnetic clusters well above $T_{\mathrm{C}}$. 
The temperature dependence of the FMR intensity follows that of the magnetization, underscoring the stability of these clusters.
We quantitatively describe both FMR and SQUID observables using atomistic spin model operating on a common set of parameters. 
 The level of agreement, achieved without tuning parameters between datasets, demonstrates the robustness and practical applicability of the approach in capturing the essential physics of spin-diluted, percolating ferromagnets.
 This study advances the understanding of percolating ferromagnetic systems, demonstrating that FMR is a key technique for probing their unique dynamic and anisotropic properties. 
 Our findings contribute to the broader exploration of dilute ferromagnets and provide new insights into percolating ferromagnetic systems, which will be relevant for spintronic opportunities.
	\end{abstract}

	

	\maketitle
	
\section{Introduction}
\label{sec:Introduction}

	Achieving direct spin control in optoelectronic devices would mark a significant technological breakthrough. In this context, dilute ferromagnetic semiconductors (DFS) have garnered considerable attention due to their unique combination of semiconductor functionality with spin degrees of freedom \cite{Ohno:1998_S,DietlOhnoMatsukuraEtAl2000}, opening avenues for spintronic applications \cite{Pearton:MaterSciEngR_2003}. Among these, Ga$_{1-x}$Mn$_x$N has attracted particular interest due to the robustness of the wide-bandgap GaN host, which makes it suitable for harsh environments.

	Despite numerous efforts \cite{Dietl:2015_RMP}, achieving room-temperature ferromagnetism (FM) in DFS remains a major challenge. In Mn-doped GaN, this difficulty arises from the insulating character of both GaN:Mn \cite{Iwinska:2019_JJAP} and (Ga,Mn)N \cite{Bonanni:2011_PRB,Yamamoto:2013_JJAP,Janicki:2017_SR}, which precludes carrier-mediated magnetic coupling. In the wurtzite GaN lattice, Mn$^{3+}$ ions ($d^4$ configuration) couple primarily via short-range ferromagnetic superexchange \cite{Blinowski:1996_PRB,Bonanni:2011_PRB,Sawicki:2012_PRB}, with interactions persisting over many coordination shells \cite{Sliwa:2024_PRB}. This long-range nature, combined with the geometric increase in Mn neighbors at greater distances, results in an effective coupling that decays approximately as $\exp(-r/b)$ with $b \approx 1.1$~\AA\, \cite{Bonanni:2021_HB}. Consequently, a percolative ferromagnetic state emerges even at low Mn concentrations \cite{Sarigiannidou:2006_PRB,Sawicki:2012_PRB,Kunert:2012_APL,Stefanowicz:2013_PRB,Gas:2018_JALCOM}, in contrast to the hole-mediated long-range FM of (Ga,Mn)As \cite{Sawicki:2006_JMMM,Jungwirth:2005_PRB(73),Jungwirth:2005_PRB(72)}.

	The percolative nature of FM in (Ga,Mn)N gives rise to a heterogeneous ensemble of magnetic clusters, both finite and infinite (macrospins), whose contributions depend on temperature and the characteristic measurement time.  
Interestingly, this phenomenon was also observed in (Ga,Mn)As, where, due to various technological factors \cite{Sadowski:2017_Nanoscale,Gluba:2018_PRB} or deliberate measurement design \cite{Sawicki:2010_NP,Proselkov:2012_APL,Chen:2015_PRL}, the material can be pushed toward the localization limit of holes.		
		This specific dynamic behavior  \cite{Dietl:2008_JPSJ,Mayer:2010_PRB,Richardella:2010_S} directly impacts quantities such as Curie temperature ($T_\mathrm{C}$), which become protocol-dependent.

	Mn$^{3+}$ ions in GaN also exhibit strong single-ion magnetic anisotropy, stemming from trigonal crystal field distortions intrinsic to the wurtzite structure \cite{Gosk:2005_PRB,Das_PairAnizo_arXiv}. 
		At low Mn concentrations, this anisotropy favors easy-plane magnetization. Moreover, Jahn–Teller effects may introduce additional threefold in-plane anisotropy, which is particularly relevant for magnetization switching in the ferromagnetic state \cite{Sztenkiel:2025_CM}.

	Previous studies have established key static magnetic properties of (Ga,Mn)N \cite{Stefanowicz:2010_PRB,Stefanowicz:2013_PRB} and recent work has demonstrated the electric field control of magnetic anisotropy \cite{Sztenkiel:2016_NC,Sztenkiel:2025_CM} and the presence of spin Hall magnetoresistance in Pt/(Ga,Mn)N heterostructures \cite{Mendozarodarte:2024_APL}. Yet, dynamic properties such as spin relaxation and magnetization precession remain underexplored.
Ferromagnetic resonance (FMR) is a powerful tool to probe these dynamics \cite{Farle:1998_Rep.Prog.Phys,Liu:2003_PRB}, but has rarely been applied to (Ga,Mn)N due to the EPR silence of Mn$^{3+}$ ions in typical X-band spectrometers \cite{Wolos:2004_PRB_69,Wolos:2004_PRB_70}. However, the presence of a robust ferromagnetic state allows for FMR detection of collective excitations, offering a route to study dynamic behavior in this otherwise EPR-silent system \cite{Stefanowicz:2013_PRB,Gas_2025}.

	On the theoretical ground, there exists a range of well-established and successfully tested approaches that describe individual aspects of (Ga,Mn)N magnetism.
Full quantum mechanical approaches such as the crystal-field model (CFM) -  originally developed for Cr$^{2+}$ in II-VI semiconductors \cite{Vallin:1970_PRB,Vallin:1974_PRB,Herbich:1998_PRB} and later adapted to Mn$^{3+}$ in GaN \cite{Graf:2002_APL,Wolos:2004_PRB_69,Wolos:2004_PRB_70,Gosk:2005_PRB,Stefanowicz:2010_PRB,Sztenkiel:2016_NC,Sztenkiel:2020_NJP,Sztenkiel:2025_CM}, provide accurate descriptions of high-field magnetization
 but treats Mn ions essentially as non-interacting. 
Extending this approach to include interactions involves constructing a Hamiltonian matrix whose dimension grows exponentially with the number of Mn spins included, i.e., $25^n \times 25^n$, where $n$ is the number of nearest-neighbor spins considered in the system. 
This renders the computation intractable beyond $n\approx 4$, and omits interactions with more distant neighbors entirely \cite{Bonanni:2011_PRB,Sztenkiel:2020_NJP,Edathumkandy:2022_JMMM,Sztenkiel:2025_CM}. 
While this method can achieve good agreement with experimental data for fields above about 30~kOe \cite{Sztenkiel:2020_NJP}, it fails to capture magnetization at low fields. 
Notably, CFM predicts vanishing remnant magnetization, $M_{\mathrm{REM}}$ = 0, at any temperature, in contradiction with experimental data showing $M_{\mathrm{REM}} \simeq 20$\% of $M(H \simeq 50~\mathrm{kOe})$  at $T=2$~K \cite{Stefanowicz:2013_PRB,Gas:2018_JALCOM} which gradually decreases with temperature and disappears at $T_{\mathrm{C}}$. 
 Monte Carlo simulations offer insight into $T_\mathrm{C}$ \cite{Simserides:2014_EPJ-WoC},  outperforming mean-field approaches based on \textsl{ab initio} inputs which tend to overestimate  $T_{\mathrm{C}}$ by up to an order of magnitude  \cite{Sato:2010_RMP},  but do not capture FMR dynamics. 
 To address this, we employ a classical atomistic spin model based on the stochastic Landau–Lifshitz–Gilbert (sLLG) formalism \cite{Evans:2014_JOP,Edathumkandy:2022_JMMM,Sztenkiel:2025_CM},  which allows us to capture both equilibrium magnetic properties and precessional dynamics, thus paving the way toward a unified simulation of both static and dynamic responses.

	In this work, we combine FMR and superconducting quantum interference device (SQUID) magnetometry to investigate a Ga$_{1-x}$Mn$_x$N layer with $x = 7.9$\%. We analyze the temperature and angular dependence of the resonance field and linewidth, extract anisotropy constants, and evaluate spin relaxation mechanisms. Importantly, we demonstrate that both FMR and SQUID data can be consistently described using a single set of parameters within our atomistic spin model, thus bridging the gap between static and dynamic magnetic characterizations of this spin-diluted percolating system.

	This paper is organized as follows. 
In Section~\ref{sec:Methods}, we describe the experimental procedures, sample preparation, and introduce the atomistic spin simulations used to model both static and dynamic magnetic properties of (Ga,Mn)N. 
Section~\ref{sec:ResultsDiscussion} presents our key findings, starting from the static magnetization data and Curie temperature analysis, followed by a detailed temperature- and angle-resolved FMR study. The section also includes a discussion of magnetocrystalline anisotropy, spin relaxation mechanisms, and a comparison between experiment and sLLG simulations. 
Finally, Section~\ref{sec:Summary} summarizes the main conclusions of the study.

	\section{Methods}
	\label{sec:Methods}
	
	\subsection {Experiment}
	
		\subsubsection {Samples}
	
	(Ga,Mn)N layers with various thicknesses up to about  1~$\mu$m are grown by the molecular beam epitaxy (MBE) technique on fully relaxed GaN templates (about 2$\mu$m thick) deposited by MOVPE on $c$-plane 0.3 - 0.4~mm thick Al$_2$O$_3$ substrates, following the specifications elaborated previously to  obtain single phase epitaxial layers without traceable amounts of other magnetic phases \cite{Kunert:2012_APL,Gas:2018_JALCOM}.
	The surface quality of the layers was monitored \textsl{in situ} during growth using reflection high-energy electron diffraction. 
	High quality and single phase of the material are confirmed  by high-resolution X-ray diffraction.

			\subsubsection {Experimental procedures}
	
	Magnetic measurements are performed in a Quantum Design superconducting quantum interference device (SQUID) MPMS-XL magnetometer with a low field option to measure magnetic properties at magnetic fields up to 70~kOe and temperatures between 2 and 400~K.
	Strict protocols are followed to avoid artifacts and limitations of volume SQUID magnetometry of thin films deposited on bulky substrates  \cite{Sawicki:2011_SST,Pereira:2011_JPDAP,Gas:2019_MST}.
	Measurements of a bare GaN template substrate serve as a  reference signal to eliminate the interference of disturbing low-temperature magnetic anisotropy of commercial sapphire substrates \cite{Gas:2022_Materials}. 
	The remaining relevant details are given in Appendix~\ref{App:Mgn}.

	Ferromagnetic resonance experiments are conducted with use of a BRUKER EMX plus spectrometer operating at 9.5 GHz ($h\nu =39$~$\mu$eV)  and equipped with an Oxford Instruments 910 Cryostat enabling temperature control from 3 to 300 K and 360 deg sample rotation with respect to the static external magnetic field $\mathbf{H}$. 
	The  recorded FMR signal is the magnetic field derivative of the absorbed microwave power that is proportional to the imaginary part of the AC magnetic susceptibility $\chi''$. 
	

	\begin{figure*}[htb]	
		\centering
		\includegraphics[width=0.98\textwidth]{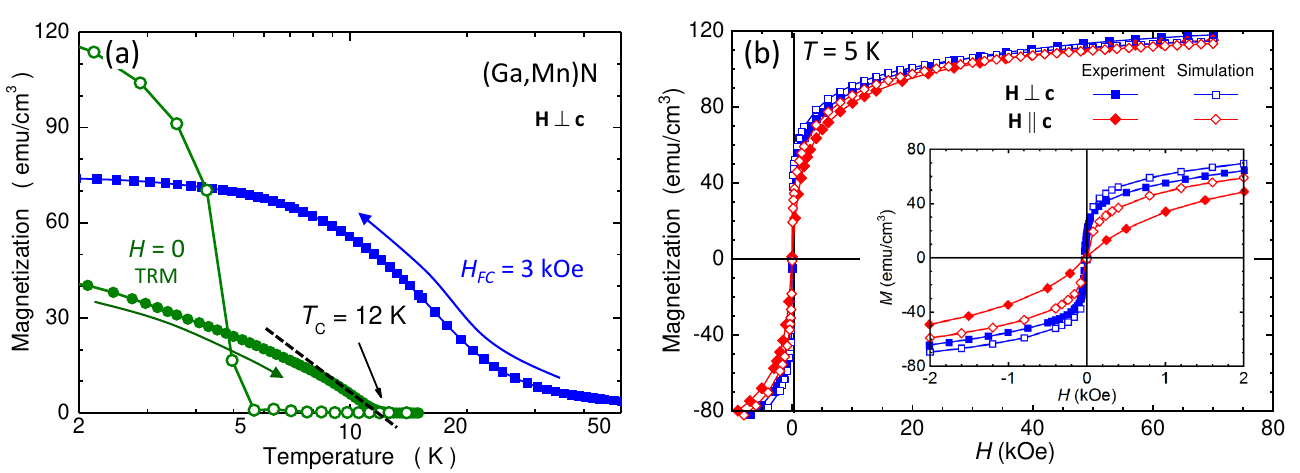}
		\caption{(Color online) Comparison of experimental (solid symbols) and simulated with the atomistic spin model (open symbols) magnetization curves for (Ga,Mn)N film with $x_{\mathrm{Mn}}=7.9$\%.
(a) Temperature dependence of the thermo-remnant magnetization (TRM) for the in-plane orientation of $H$ (full green circles). The initial field-cooled (FC) magnetization measured during cooling at $H_{\mathrm{FC}} = 3$~kOe is shown as solid blue squares. The temperature at which TRM vanishes defines the Curie temperature, $T_{\mathrm{C}}$, of the film. Open circles represent results from atomistic spin model and Monte Carlo simulations. (b) Magnetization as a function of magnetic field, $H$, measured at $T=5$~K for two field orientations: perpendicular to the $c$-axis (full squares) and along the $c$-axis (full diamonds), and results of the atomistic spin model added as matched open symbols. The inset provides an expanded view of the low-field region.
		}
		\label{Fig:REM}
	\end{figure*}
	
		\subsection {Numerical simulations}
		
	The experimental data are modeled using an atomistic spin model \cite{Evans:2014_JOP} combined with the stochastic Landau-Lifshitz-Gilbert (sLLG) equation \cite{Edathumkandy:2022_JMMM, Sztenkiel:2025_CM}. 
	Simulations are performed on a system consisting of 9860 randomly distributed Mn spins in a wurtzite lattice, with in-plane periodic boundary conditions to model a large thin film. 
	The classical spin Hamiltonian accounts for Zeeman energy, magnetocrystalline anisotropy, and ferromagnetic superexchange interactions between Mn ions taken exactly up to the 14$^{th}$ coordination sphere \cite{Bonanni:2021_HB}. 
	The spin dynamics are governed by the sLLG equation, and the magnetization and FMR response are computed as a function of external magnetic field and temperature. 
	A Monte Carlo method is employed to simulate thermo-remnant magnetization. \cite{Zhang:2021_CMS,Prudnikov:2016_JP}.
	 
	Of particular importance in this study is an attempt to model FMR measurements. 
To this end the simulation box is numerically subjected to an AC magnetic field  \cite{Usadel:2006_PRB}, and the resonance field and line-width are extracted from the simulation results.

Details of the simulation methodology, including the full Hamiltonian, equations of motion, parameter values, and numerical integration scheme are provided in Appendices~\ref{App:Simul} and~\ref{App:Integr}, while additional discussion of the model assumptions and limitations is given in Appendix~\ref{App:Limitations}.

	\section{RESULTS AND DISCUSSIONS}
\label{sec:ResultsDiscussion} 
	
	\subsection{Magnetic characterization}

The Curie temperature, $T_{\mathrm{C}}$, is a key parameter of ferromagnetic materials \cite{Fabian:2013_GGG}, but its determination remains challenging—especially in magnetically dilute systems, where critical features are often smeared over a broad temperature range. 
In this study, we employ thermoremnant magnetization (TRM) measurements as a robust method for determining $T_{\mathrm{C}}$ in dilute ferromagnetic semiconductors, regardless of the FM interaction range \cite{Stefanowicz:2013_PRB,Wang:2013_PRB,Gas:2018_JALCOM,Gas:2021_JMMM_Cr}.

To acquire TRM data the sample is initially field-cooled in a magnetic field of $H_{FC} = 3$~kOe [blue squares in Fig.~\ref{Fig:REM}~(a)].
After quenching the field at $T=2$~K the values of the remnant magnetisation $M_{\mathrm{TRM}}$ is recorded during warming [green bullets in Fig.~\ref{Fig:REM}~(a)]. 
The temperature at which $M_{\mathrm{TRM}}$ vanishes indicates  $T_{\mathrm{C}}=12$~K in this film.
More in-depth experimental considerations pertinent to this technique are outlined in Appendix~\ref{App:Mgn}.

	Exemplary magnetization curves, $M(H)$, obtained for the same sample at $T= 5$~K~$< T_{\mathrm{C}}$ are shown in Fig.~\ref*{Fig:REM}~(b) as solid symbols. 
		Despite being in the FM state, (Ga,Mn)N exhibits markedly rounded $M(H)$ curves at low fields \cite{Sarigiannidou:2006_PRB,Kunert:2012_APL,Stefanowicz:2013_PRB,Gas:2018_JALCOM,Kalbarczyk:2019_JALCOM,Sztenkiel:2020_NJP,Sztenkiel:2025_CM}. 
	This behavior stems from magnetic dilution and the short-range nature of superexchange interactions.
	
On a microscopic level, this leads to a broad distribution of ferromagnetically coupled clusters, where hysteresis at low fields involves only a fraction of spins. 
At this dilution level, (Ga,Mn)N does not form a uniform FM phase below $T_{\mathrm{C}}$. 
Moreover, $M(H)$ does not fully saturate even at 70 kOe, reflecting a significant contribution of the weakly saturating orbital moment of Mn$^{3+}$ ions \cite{Sztenkiel:2023_JMMM}.

The atomistic spin simulations corresponding to the $M_{\mathrm{TRM}}(T)$ and $M(H)$ data in Fig.~\ref{Fig:REM} are shown as open symbols. 
Both regimes are simulated using a single parameter set, initially optimized to match the experimental $M(H)$ curves. 
As expected, agreement is very good in the high-field region, while discrepancies arise at low fields (see inset of Fig.~\ref{Fig:REM}~(b)) and in the $M_{\mathrm{TRM}}(T)$ simulation. 
In both cases, the model slightly overestimates magnetization at low fields.

 These deviations stem from three limitations of the model, discussed in Appendix~\ref{App:Limitations}: (i) omission of pairwise Mn--Mn anisotropy effects \cite{Das_PairAnizo_arXiv}; (ii) approximation of Jahn--Teller anisotropy with a cubic term; and (iii) assumption of perfectly random Mn distribution. 
 All these factors affect the local anisotropy landscape, especially relevant when Zeeman energy is insufficient to overcome anisotropy barriers.

Achieving a Curie temperature within a factor of two of experiment is thus a reasonable result. 
We note that doubling the exchange integral would reproduce the experimental $T_{\mathrm{C}}$, but also lead to unrealistic rectangular $M(H)$ loops and overestimated coercive fields — an outcome clearly at odds with the experiment.

Despite its simplicity, the model captures key experimental features and provides a sound basis for simulating the FMR response, as discussed in Sections~\ref{subsec:FMR_MgnCrystAni} and \ref{subsec:FMR_line_width}.

	\subsection{Temperature dependence of the FMR}
	
	\begin{figure}
		\subfigure{
			\includegraphics[width=0.42\textwidth]{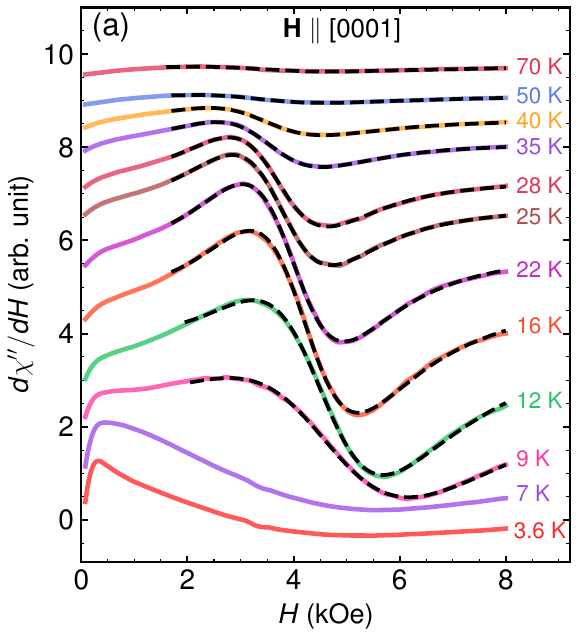}
		}
		\subfigure{
			\includegraphics[width=0.43\textwidth]{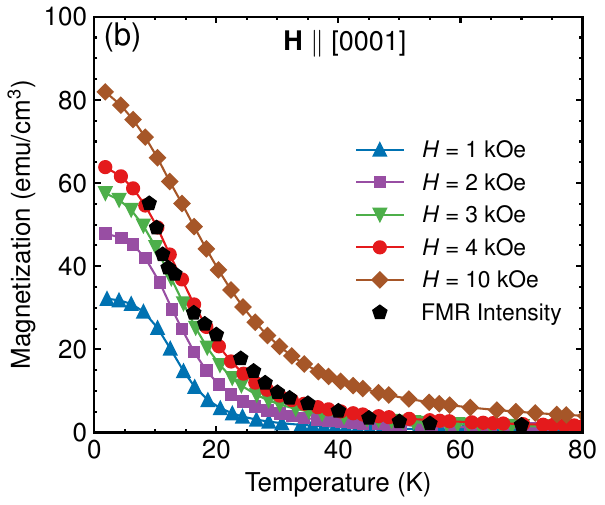}
		}
		\caption{\label{Fig:FMR_T}(Color online)~(a)~Selected, temperature dependent ferromagnetic resonance spectra measured with magnetic field $\mathbf{H}$ applied along the out-of-plane, [0001] crystallographic axis of the (Ga,Mn)N film. The spectra are shifted vertically for clarity. Black dashed lines are fits of Eq.~\ref{Eq:Lorentzian}. 
			(b)~Temperature dependence of the ferromagnetic moment determined by double integration of the FMR signals (pentagons), compared with the magnetization of the (Ga,Mn)N film measured in the same geometry at several magnetic fields $H$ (indicated in the graph).
		}
	\end{figure}

	Fig.~\ref{Fig:FMR_T}~(a) shows the temperature dependence of the FMR spectra for $\mathbf{H}\parallel[0001]$. 
	This hard-axis orientation minimizes the overlap between the FMR signal and the cyclotron resonance (CR) background from the conducting GaN template. 
	At low temperatures ($T < 7$~K), only a strong CR and a weak $g=2$ paramagnetic signal are detected. 
	Ferromagnetic features emerge near $T = 9$~K and persist up to 70~K, with their integrated intensity decreasing with increasing temperature. 
	This behavior contrasts sharply with that observed in other DFSs, such as (Ga,Mn)As \cite{Liu:2003_PRB,Liu:2006_JPCM,Zhou:2009_PRB}, (Ga,Mn)P \cite{Bihler:2007_PRB}, and (Pb,Mn)(Te,Se) \cite{Story:1993_PRB}, where FMR signals are readily observed well below $T_{\mathrm{C}}$.

We attribute the absence of detectable FMR signals below $T_{\mathrm{C}}$ to strong homogeneous broadening, likely caused by an enhanced magnetization relaxation rate, discussed in Section~\ref{subsec:FMR_line_width}. 
	A plausible origin of this rapid relaxation is the high density of structural defects, particularly threading dislocations, typical for (Ga,Mn)N films grown on foreign substrates \cite{Hsu:2001_APL,Kalbarczyk:2019_JALCOM}.

To compare the FMR and SQUID results, we fit the spectra using the derivative of a Lorentzian function 
	\begin{equation}\label{Eq:Lorentzian}
\frac{A}{\pi} \frac{d}{dH} \left[ \frac{\Delta H}{(H - H_R)^2 + \Delta H^2}
+ \frac{\Delta H}{(H + H_R)^2 + \Delta H^2} \right],
\end{equation}
to extract the amplitude $A$, resonance field $H_R$, and the half width at half maximum (HWHM) $\Delta H$ at each temperature. 
The fits are denoted by black dashed lines in Fig.~\ref{Fig:FMR_T}~(a).
The FMR intensity, $I$, is then calculated as $I=A \times \Delta H$ and is directly proportional to the magnitude of the ferromagnetic moment. 
While absolute magnetization values cannot be determined due to magnetocrystalline anisotropy, we normalize $I(T)$ to the SQUID magnetization at 3~kOe and 55~K. 
As shown in Fig.~\ref{Fig:FMR_T}~(b), $I(T)$ closely follows $M(T)$ between 3–4~kOe. 
The persistence of the FMR signal up to 70~K likely reflects both the stabilizing effect of the applied field \cite{Sasaki:2002_JAP,Stone:2006_APL} and the presence of non-percolating FM clusters above $T_{\mathrm{C}}$.

	\begin{figure}
		\centering
		\includegraphics[width=0.45\textwidth]{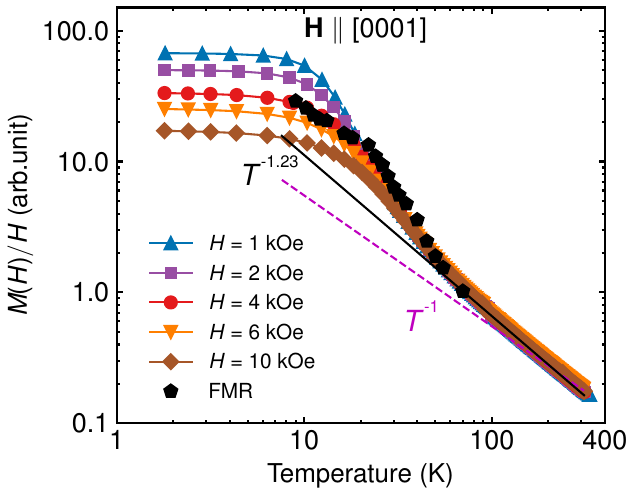}
		\caption{ \label{Fig:LogLog}
			(Color online)~Temperature $T$ dependence of the magnetic susceptibility, $\chi(T) = M(T)/H$, measured for the (Ga,Mn)N film in magnetic fields between  1 and 10~kOe, and plotted in a double logarithmic scale (small colored symbols). Black pentagons represent the temperature dependence of the FMR intensity normalized by the resonance field  $I/H_R$. The dashed magenta line indicates the Curie law behavior, i.e. the proportionality to $1/T$, whereas the solid black line follows $\chi(T) \propto T^{-1.23}$ at high temperatures, as expected for a random paramagnet with ferromagnetic correlations.
		}
	\end{figure}

	\begin{figure*}[t]
		\subfigure{
			\raisebox{0.54cm}{
				\includegraphics[width=0.272\textwidth]{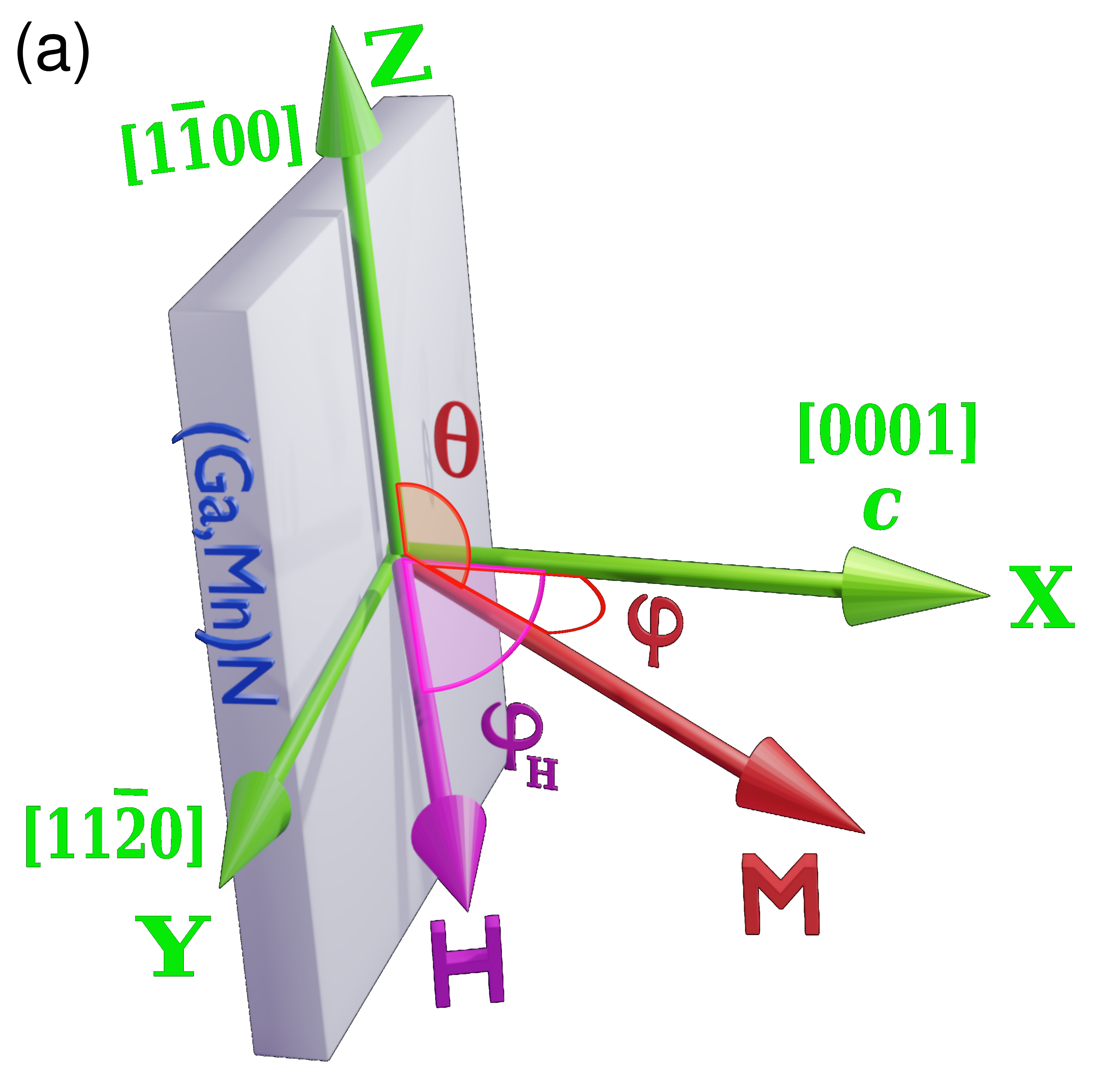}
			}
		}
		\subfigure{
			\includegraphics[width=0.333\textwidth]{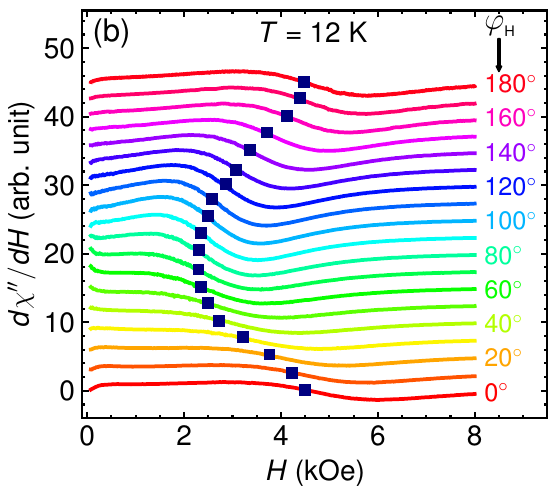}
		}
		\subfigure{
			\includegraphics[width=0.333\textwidth]{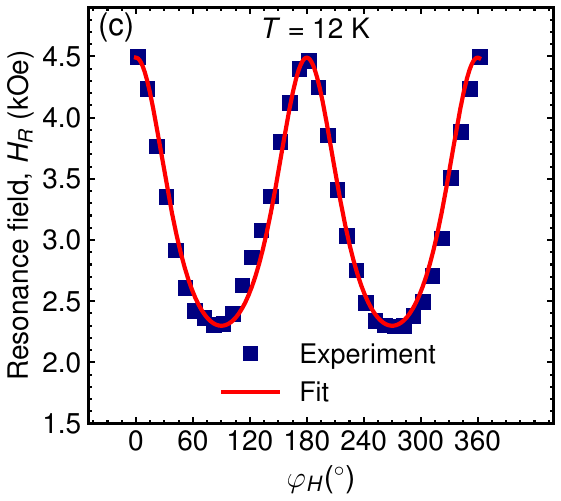}
		}
		\caption{\label{fig:angular_dependent_FMR_signal}(Color online)~(a) Scheme of the FMR measurements geometry emphasizing the coordinate system used. The applied dc magnetic field vector $\mathbf{H}$ lies in the $(1\bar{1}00)$ plane and the azimuthal angle $\varphi_H$ is counted from the [0001] (X) axis. The orientation of the magnetization vector $\mathbf{M}$ is given by the azimuthal $\varphi$ and polar $\theta$ angles, the latter counted from the $[1\bar{1}00]$ (Z) axis. In principle, $\mathbf{H}$ and $\mathbf{M}$ are not collinear. (b)~Selected, angularly dependent ferromagnetic resonance spectra recorded with the magnetic field rotated in the $(1\bar{1}00)$ plane at $T=12$~K. Symbols indicate resonance field values. (c) Angular dependence of the resonance fields $H_R$ for the magnetic field $\textbf{H}$ rotated in the $(1\bar{1}00)$ plane at $T=12$~K. Symbols represent experimental data, the solid line is a fit.}
	\end{figure*}

	To explore this further, we replot the FMR and SQUID data in a double-logarithmic scale [Fig.~\ref{Fig:LogLog}]. 
	In disordered systems with varying exchange interactions, magnetic susceptibility $\chi(T) = M(T)/H$ follows a power-law: $\chi(T) \propto T^{-\alpha}$ \cite{Bhatt:1986_PS,Anderson:1986_PRB}. 
	For noninteracting moments, $\alpha = 1$ (Curie law). 
	Antiferromagnetic interactions yield $\alpha < 1$ \cite{Bhatt:1982_PRL,Dietl:1987_JJAP,Sawicki:2013_PRB}, while ferromagnetic ones give $\alpha > 1$, as observed in (Ga,Mn)N \cite{Stefanowicz:2013_PRB}. 
	Our data follow $\alpha = 1.23$ for $T > 80$~K, independent of $H$ in the 1–10~kOe range (indicated by the solid black line in Fig.~\ref{Fig:LogLog}).

Below $T_f = 70$~K, $\chi(T)$ curves deviate upward from the $T^{-1.23}$ trend and begin to fan out below 30~K, without clear critical behavior near $T_{\mathrm{C}}$. 
Notably, the FMR signal also emerges at $T_f$, where $\chi(T)$ begins to exceed the expected quasi-paramagnetic $T^{-1.23}$ response. 
This coincidence suggests that $T_f$ marks the onset of coherent microwave absorption by FM-coupled macrospins.

	The nature and size of these entities are of key interest. 
	Quantum mechanical calculations on Ga$_{0.97}$Mn$^{3+}_{0.03}$N including up to fourth-order Mn complexes showed no deviation from paramagnetic behavior \cite{Sztenkiel:2020_NJP}, implying that macrospin puddles involve $N \ggg 4$ Mn ions—consistent with a percolation picture.

Although Mn$^{3+}$ ions in GaN tend to cluster \cite{Gonzales:2011_PRB}, years of materials optimization \cite{Sarigiannidou:2006_PRB,Stefanowicz:2010_PRBa,Kunert:2012_APL,Gas:2018_JALCOM} have led to nearly homogeneous Mn distribution. Residual macroscopic inhomogeneities exhibit a narrow variance $\Delta x \simeq 0.2$\% \cite{Stefanowicz:2013_PRB}, insufficient to indicate phase separation, such as spinodal decomposition \cite{Bonanni:2010_CSR}, known to cause spurious ferromagnetism in related systems. Resolving the fine structure of Mn clustering would require atomically resolved techniques like Atom Probe Tomography, which lie beyond the scope of this work.

	\begin{figure*}[htb]
		\centering
		\begin{minipage}{\textwidth}
			\centering
			\includegraphics[width=.35\textwidth]{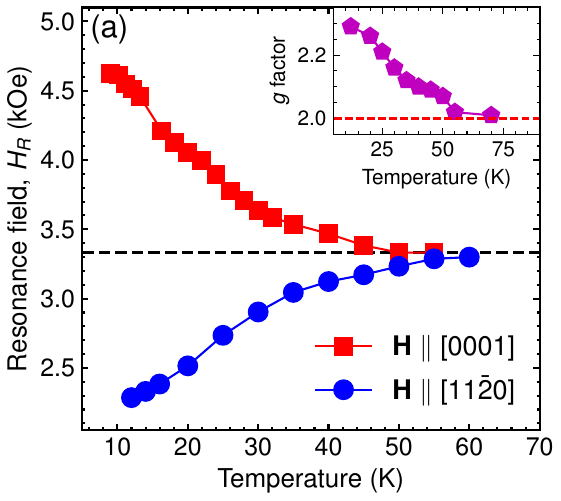}\quad 
			\includegraphics[width=.35\textwidth]{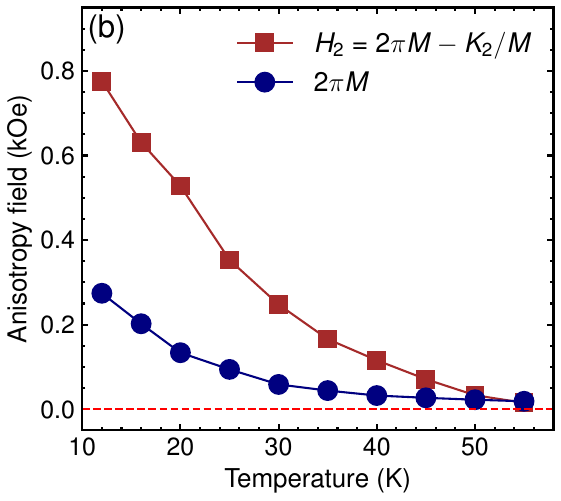}
			\vspace{0.25cm} 
			
			\includegraphics[width=.35\textwidth]{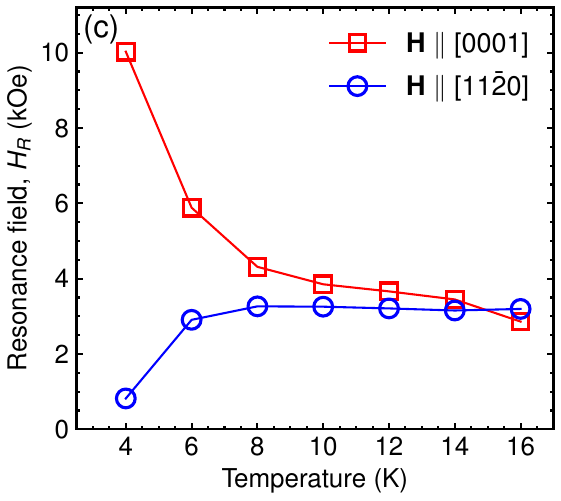}\quad 
			\includegraphics[width=.356\textwidth]{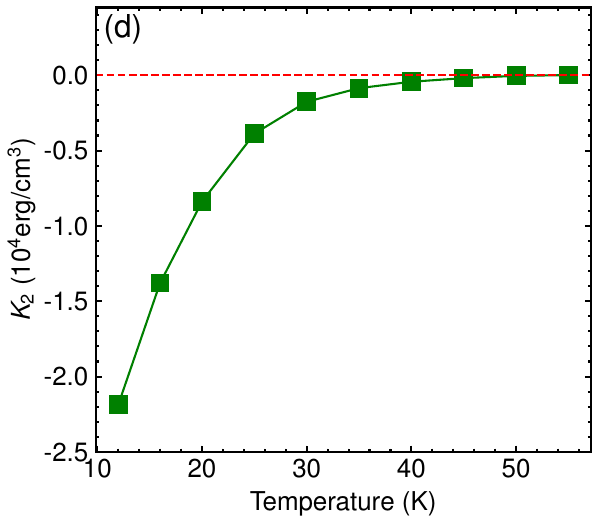}
		\end{minipage}
		\caption{\label{fig:resonance field_temperature}~(Color online)~(a)~Temperature dependence of the resonance fields $H_R$ for two magnetic field orientations: $\mathbf{H}\parallel[0001]$ (squares) and $\mathbf{H}\parallel[11\bar{2}0]$ (circles). The dashed line marks the field corresponding to $g=2$. The inset shows the temperature dependence of the effective g-factor. 
			(b)~Temperature dependence of the anisotropy field $H_2$ and the demagnetization field $2\pi M$ (squares and circles, respectively). The latter is determined using the magnetization data shown in the inset of Fig.~\ref{Fig:FMR_T} for $H=4$~kOe. 
			(c)~Simulation results depicting the temperature dependence of $H_R$ for two magnetic field orientations: $\mathbf{H}\parallel[0001]$ (open squares) and $\mathbf{H}\parallel[11\bar{2}0]$ (open circles). 
			(d)~Temperature dependence of the uniaxial magnetocrystalline anisotropy energy $K_2$, obtained from (b). The solid line is a guide for the eye.}
	\end{figure*}

	\subsection{Magnetocrystalline anisotropy}
		\label{subsec:FMR_MgnCrystAni}  
	
	Ferromagnetic resonance is the easiest experimental technique to investigate magnetocrystalline anisotropy. 
	For this purpose we have chosen to rotate the external magnetic field in the (1$\bar{1}$00) plane, which contains all the relevant crystallographic directions, including the $\langle 100 \rangle$ cubic axes along which static Jahn-Teller distortions occur \cite{Wolos:2004_PRB_b}. 
	However, the low temperature line broadening and the background cyclotron resonance of the GaN template precludes angular dependent investigations below 12~K. 
	The experimental geometry is shown schematically in Fig.~\ref{fig:angular_dependent_FMR_signal}~(a). 
	Selected FMR spectra recorded for different orientations of the applied magnetic field are presented in Fig.~\ref{fig:angular_dependent_FMR_signal}~(b), while the angular dependence of the determined resonance fields is given in Fig.~\ref{fig:angular_dependent_FMR_signal}~(c) (symbols). 
	The solid line in Fig.~\ref{fig:angular_dependent_FMR_signal}~(c) is calculated with the use of the Smit-Beljers equation \cite{Smit:1955_PRR}:
	\begin{eqnarray} \label{smit} 
		h\nu = \frac{g\mu_B}{M \sin\theta_{eq}}
		\sqrt{ \frac{\partial^2 F}{\partial \theta^2} \frac{\partial^2 F}{\partial \varphi^2}
			- \left( \frac{\partial^2 F}{\partial \theta \partial \varphi}
			\right)^2 }
		\bigg|_{\theta_{eq}, \varphi_{eq},}
	\end{eqnarray}
	where $\nu =$ 9.5 GHz is the uniform mode resonance frequency, $g$ is the effective spectroscopic splitting factor, $\mu_B$ is the Bohr magneton, $\theta$ and $\varphi$ are the polar and azimuthal angles of the magnetization vector, $\textbf{M}$, defined in Fig.~\ref{fig:angular_dependent_FMR_signal}~(a),  and  $F$ is the anisotropic part of the magnetic free energy density. 
	The equilibrium angles of $\textbf{M}$ ($\theta_{eq},\varphi_{eq}$) are determined by minimizing $F$. 
	In the coordinate system of Fig.~\ref{fig:angular_dependent_FMR_signal}~(a) $F$ is given by:
	\begin{multline}\label{energy}
		F=-MH \sin\theta \cos(\varphi-\varphi_H)+\\(2\pi M^2-K_2)\sin^2 \theta \cos^2 \varphi.
	\end{multline}
	Here, the first term describes the Zeeman energy, whereas the second term is a combination of the thin film demagnetization energy $(2\pi M^2)$ and the uniaxial magnetocrystalline anisotropy energy $(K_2)$ along the [0001] axis. 
	We adhere to the standard convention  that the sign of the uniaxial anisotropy is opposite to that defined in the electron paramagnetic resonance Hamiltonian \cite{Farle:1998_Rep.Prog.Phys}. 
	Higher order terms in $F$ are omitted.
	Since $\frac{\partial F}{\partial \theta}=0$ for $\theta_{eq}=90^{\circ}$  the magnetization vector also lies in the (1$\bar{1}$00) rotation plane. 
	Defining an effective uniaxial anisotropy field $H_2$  as: 
		\begin{equation}\label{Eq:H2}
		H_2 = 2\pi M -\frac{K_2}{M}
	\end{equation}
	we obtain the following equations:   
	\begin{multline}\label{smit_condition}
		\left(\frac{h\nu}{g\mu_B}\right)^2 =\left\{H_R \cos(\varphi-\varphi_H)-2H_{2} \cos^2\varphi \right\} \times\\
		\left\{H_R \cos(\varphi-\varphi_H)-2H_{2} \cos 2\varphi \right\},
	\end{multline}
	\begin{equation}\label{Hsin}
		H_R \sin(\varphi-\varphi_H)=H_{2} \sin 2\varphi,
	\end{equation}
	which can be solved analytically.
	The relation between $\varphi_H$ and $\varphi$ is given by:
	\begin{equation}\label{phi_H}
		\varphi_H = \varphi - \arcsin\left( \frac{ H_2 \sin 2\varphi}{H_R} \right).
	\end{equation}
		The best fit of  Eqs:\eqref{smit_condition}-\eqref{phi_H} to the experimental data is shown by solid line in Fig.~\ref{fig:angular_dependent_FMR_signal}~(c).
	It  yields $H_2=0.77$~kOe and $g=2.29$. 
	The uniaxial anisotropy energy density, determined by subtracting the magnetization value obtained from SQUID measurements at 4~kOe and $T=12$~K, is $K_2 = -2.1 \times 10^4 \, \text{erg}/\text{cm}^3$.

	The temperature dependence of the uniaxial anisotropy is obtained from FMR signals at two orientations of the applied magnetic field: $\mathbf{H}\parallel[0001]$ ($\varphi_{_{{H}}}=0^{\circ}$) and $\mathbf{H}\parallel[11\bar{2}0]$ ($\varphi_{_{{H}}}=90^{\circ}$). 
	The temperature dependencies of $H_R$ for these two field orientations are shown in Fig.~\ref{fig:resonance field_temperature}~(a). 
	With increasing temperature the resonance fields  monotonically approach the value corresponding to $g=2$, characteristic of non-interacting Mn$^{3+}$ ions, as shown by the temperature dependence of the effective $g$-factor in the inset. 
	This finding is consistent with the magnetic $M(H)/H$ vs.~$T$ data (Fig.~\ref{Fig:LogLog}), where  deviations from the high-temperature trend $T^{-1.23}$ occur at approximately the same temperature at which $g$-factor starts to deviate from 2, indicating the onset of detectable ferromagnetic coupling among some  Mn$^{3+}$ ions.

	The temperature dependence of the effective anisotropy fields $H_2$ determined from the data presented in Fig.~\ref{fig:resonance field_temperature}~(a) is shown in Fig.~\ref{fig:resonance field_temperature}~(b), together with that of demagnetization fields $2\pi M$ obtained from SQUID measurements at $H=4$~kOe (Fig.~\ref{Fig:FMR_T}~(b)). These two dependencies allow us to approximate the temperature behavior of the uniaxial magnetocrystalline energy density $K_2$ which is shown in Fig.~\ref{fig:resonance field_temperature}~(d). 
	As can be seen, $K_2$ is negative in accordance with the fact that the [0001] crystallographic axis ($c$ axis) of GaN is the hard magnetization direction in (Ga,Mn)N. 
	With increasing temperature the absolute value of $K_2$ decreases monotonically to zero. 
	
	Figure~\ref{fig:resonance field_temperature}~(c) illustrates the temperature dependence of the resonance field $H_R(T)$ obtained from numerical simulations. 
	As in the experimental data, we analyze two configurations: $\textbf{H}$ applied along the out-of-plane [0001] direction and the in-plane $[11\bar{2}0]$ direction.
	The simulation results qualitatively reproduce the experimental trends shown in Fig.~\ref{fig:resonance field_temperature}~(c), albeit in about four times narrower temperature range. 
	However, the same pronounced anisotropy in $H_R$ is observed at low temperatures between the two directions, with the out-of-plane configuration exhibiting significantly larger resonance fields. 
	This anisotropy gradually diminishes with increasing temperature, and $H_R$ converges to approximately the same
	value as observed in the experiment, about 3.3~kOe.

	The low value of the uniaxial anisotropy constant $K_2$ as compared to those reported for various ferromagnetic transition metals showing room temperature ferromagnetism (e.g.~Ref.\cite{Farle:1998_Rep.Prog.Phys} and references therein) is easily explained bearing in mind the rather moderate magnetic moment resulting both from the low Mn concentration as well as the partial quenching of the spin–orbit interaction (which is the main source of anisotropy in non-metallic ferromagnets) by the Jahn-Teller effect \cite{Wolos:2004_PRB_b}. 
	As the 4-th order terms in the magnetic free energy density are usually an order of magnitude weaker than the uniaxial terms, the lack of such anisotropy terms in the experimental data is not unexpected.
	
	Interestingly, no influence of the static Jahn-Teller distortion on the $H_R(\varphi)$ dependence is observed, contrary to expectations. 
	Specifically, no broadening of the FMR signals is detected along the  $\langle 100 \rangle$ cubic axes directions (detailed in the next Section). 
	This result suggests a transition from a static Jahn-Teller effect, previously reported at 2~K  \cite{Wolos:2004_PRB_b}, to a dynamic regime at $T \simeq 12$~K, where cubic distortion directions are averaged. 
	This observation is intriguing and warrants further investigation.
	
	\begin{figure}[t]
		\includegraphics[width=0.8\linewidth]{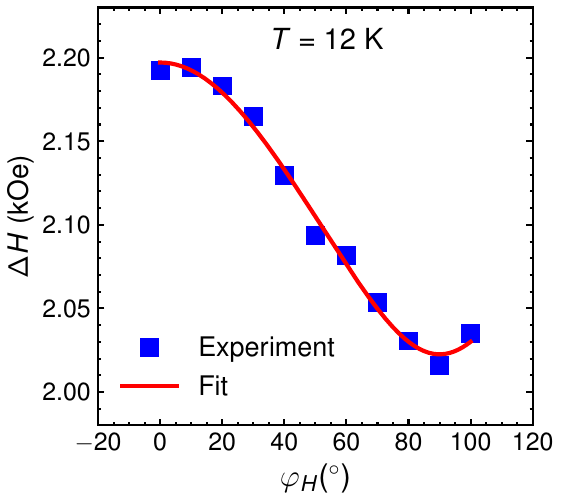}
		\caption{\label{fig:linewidth_temperature}(Color online)~Angular dependence of the linewidth $\Delta H$ measured at 12~K with the applied magnetic field $\textbf{H}$ rotated in the $(1\bar{1}00)$ crystal plane (symbols). The line is a fit obtained using the sum of Eqs.~\eqref{eq:homo2} and \eqref{eq:inhomo} for $\Delta H_{homo} = 2.07$~kOe and $\Delta H_2=63$~Oe.}
	\end{figure}
	
	\subsection{FMR linewidth and relaxation in (Ga,Mn)N }
	\label{subsec:FMR_line_width}  

	The investigation of the FMR linewidth ($\Delta H$) can give valuable information on the relaxation rate of the magnetization as well as potential inhomogeneities of the studied system. $\Delta H$ contains the contributions originating from the intrinsic damping of the magnetization precession, 	$\Delta H_{homo}$, and the contribution related to magnetic and/or structural inhomogeneities in the system. 
 
In the $(1\bar{1}00)$ rotation plane 	$\Delta H_{homo}$ simplifies to \cite{Platow:1998_PRB}:
	\begin{equation}\label{eq:homo2}
		\Delta H_{homo}(\theta,\varphi) = \left( \frac{\hbar}{g \mu_B} \right)^2 \frac{G \omega}{M \cos(\varphi - \varphi_H),}
	\end{equation}
		where $G$ is the Gilbert damping parameter.
	One should, therefore, expect minimum linewidths only for two magnetic field directions, since $\varphi_H - \varphi = 0$ only for in-plane and out-of-plane orientations of $\mathbf{H}$. Instead, the angular dependence of the linewidth measured at 12~K and shown in Fig.~\ref{fig:linewidth_temperature} resembles that of $H_R$   [Fig.~\ref{fig:angular_dependent_FMR_signal}~(c)], with maximum and minimum $\Delta H$ values at $\mathbf{H}\parallel[0001]$ and $\mathbf{H}\parallel[11\bar{2}0]$, respectively. 
	Such a dependence is expected for inhomogeneous line broadening, arising from a distribution of magnetic moments and anisotropy fields within the sample, which can be expressed as \cite{Chappert:1986_PRB}: 
	\begin{equation}\label{eq:inhomo}
		\Delta H_{inhomo} = \Delta H_2 \frac{\partial H_R}{\partial H_2}.
	\end{equation}
	The fit of experimental data with the sum of  Eqs.\eqref{eq:homo2} and \eqref{eq:inhomo}  presented by the solid line in Fig. \ref{fig:linewidth_temperature} gives $\Delta H_{homo} = 2070$~Oe and $\Delta H_2=63$~Oe. 
	In the fit the angular dependence of $\Delta H_{homo}$ is neglected since the maximum deviation of $\varphi$ from $\varphi_H$ (at $\varphi_H = 50$ deg) leads only to 1.5\% line broadening, which falls within the experimental error.
	It is evident that the main contribution to the linewidth is due to intrinsic damping. 
	The Gilbert parameter determined from Eq.~\ref{eq:homo2} is about 4$\times 10^8~s^{-1}$ for $\textbf{H}$ along [0001] at 12 K and does not differ much from the values found for typical ferromagnetic metals \cite{Farle:1998_Rep.Prog.Phys}. 
	The reason for the untypically broad linewidth is the low value of magnetization [$M \simeq 45$~emu/cm$^3$, as indicated in Fig.~\ref*{Fig:FMR_T}~(b)], which enters in the denominator of Eq.\eqref{eq:homo2}. 
	This explains also the unusually high value of the parameter $\alpha_G = \frac{\hbar G}{g \mu_B M}$ = 0.68 as compared to typical values observed in diluted magnetic systems and metals \cite{Farle:1998_Rep.Prog.Phys,Sinova:2004_PRB,Matsuda_2006_PhysicaB}.
	\begin{figure*}[ht]
		\subfigure{
			{
				\includegraphics[width=0.313\textwidth]{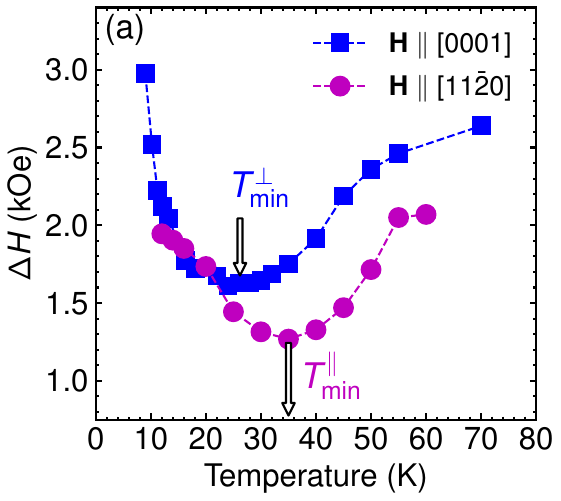} 
			}
		}
		\subfigure{
			\includegraphics[width=0.313\textwidth]{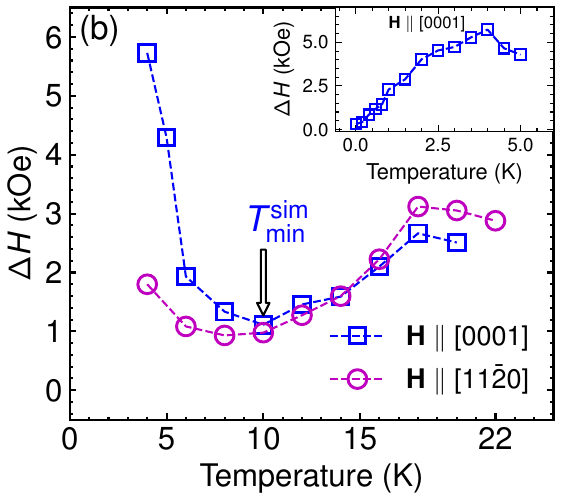}
		}
		\subfigure{
			\includegraphics[width=0.313\textwidth]{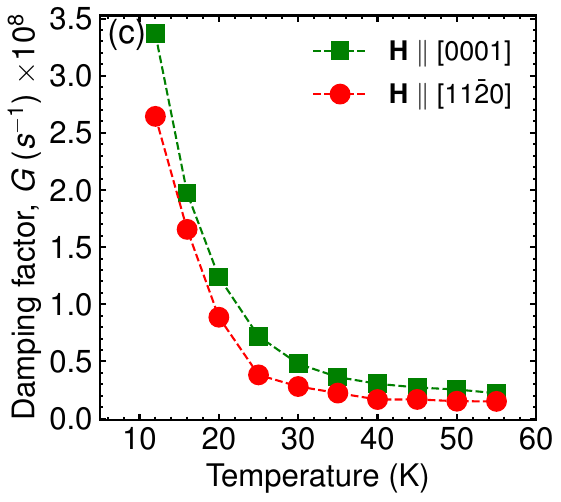}
		}
		\caption{\label{fig:linewidth_critical fluctuations}(Color online)~(a)~Measured temperature dependence of the linewidth $\Delta H$ for two magnetic field orientations: $\mathbf{H}\parallel[0001]$ (full squares) and $\mathbf{H}\parallel[11\bar{2}0]$ (full circles). The temperatures at which $\Delta H$ attains its minimum values are marked as $T_\mathrm{min}^{\perp}$ and $T_\mathrm{min}^{\parallel}$, respectively. (b)~Simulated temperature dependence of the linewidth $\Delta H$ for the  magnetic field orientation $\mathbf{H}\parallel[0001]$ (open squares) and $\mathbf{H}\parallel[11\bar{2}0]$ (open circles). The inset in (b) shows the computed results in a wider temperature range for $\mathbf{H}\parallel[0001]$.~(c)~Temperature dependence of damping factor, $G$, for two magnetic field orientations: $\mathbf{H}\parallel[0001]$ (full squares) and $\mathbf{H}\parallel[11\bar{2}0]$ (full circles). Lines are guides for the eye.}		
	\end{figure*}

	The contribution of homogeneous broadening to the total linewidth at 12~K is below 6\%, despite the relatively large value of the anisotropy field ($H_2=0.77$~kOe). 
	This is attributed to the small relative variation of the anisotropy field, $\Delta H_2/H_2= 8$\%, which serves as a measure of the spatial inhomogeneity in the Mn distribution.
		This level of variation is consistent with the previously reported relative standard deviation of Mn concentration, $\Delta x / x \simeq 2-4$\%, in similar samples \cite{Stefanowicz:2013_PRB}.
	
	Fig.~\ref{fig:linewidth_critical fluctuations}~(a) illustrates the temperature dependence of $\Delta H$ measured for $\mathbf{H}$ applied along the out-of-plane [0001] and in-plane [11$\bar{2}$0] directions. 
	As shown, the linewidth narrows with increasing temperature in both configurations, reaching minimum values at approximately 26~K and 35~K, as marked in the figure. 
	This narrowing is only partly due to decreasing anisotropy fields [shown in Fig.~\ref{fig:resonance field_temperature}~(b)], but is mostly due to a reduction of the the Gilbert damping parameter $G$ with increasing temperature. 
	The temperature dependence of $G$ extracted from the experimental data is presented in Fig.~\ref{fig:linewidth_critical fluctuations}~(c) for in- and out-of-plane magnetic field directions. 
	As can be seen, the damping parameter decreases monotonically with increasing temperature and is consistently greater for the magnetic field applied along the hard axis than for the in plane orientation.   
	The anisotropy of $G$ contributes to the observed difference in the temperature dependencies of the in-plane and out-of-plane linewidths shown  in Fig.~\ref{fig:linewidth_critical fluctuations}~(a).

	The sharp increase in linewidths above the minimum temperatures, up to about 55~K, is attributed to enhanced homogeneous damping associated with the decreasing magnetization of individual clusters, as described by Eq.\eqref{eq:homo2}. 
	These clusters contain  Mn$^{3+}$ ions that remain sufficiently strongly coupled to produce a detectable FMR signal at temperatures well above 12~K, the Curie temperature of the film determined from direct magnetization measurements.

	The observed trend in $\Delta H(T)$ is typical for many ferromagnetic systems  \cite{Platow:1998_PRB, Li:1990_PRB, Kotzler:1978_PRL, Sporel:1975_SSC}. The same trend is visible in Fig.~\ref{fig:linewidth_critical fluctuations}~(b) which illustrates the temperature dependence of $\Delta H$ obtained from simulations. Similar to experimental observations a significant broadening of $\Delta H$ is noted at lower temperatures. As the temperature increases, $\Delta H$ narrows, reaching a minimum value of 1~kOe at approximately at 10~K denoted by $T_\mathrm{min}^{\mathrm{sim}}$. Upon further temperature increase, $\Delta H$ slightly broadens again, reaching a maximum at 18~K. 
	Although the obtained values differ from the experimental ones, they are comparable if presented in reduced temperature $T/T_{\text{C}}$.

	\section{Summary}
\label{sec:Summary}

In this study, we have investigated the magnetic anisotropy and relaxation dynamics in percolating ferromagnetic semiconductor Ga$_{1-x}$Mn$_{x}$N with $x \sim 8$\% using ferromagnetic resonance (FMR) and SQUID magnetometry. Our results confirm the ferromagnetic nature of the material, dominated by Mn$^{3+}$ ions, with a well-defined transition at $T_{\mathrm{C}} = 12$~K.
Although the FMR signal vanishes below $T \approx 9$~K due to excessive line broadening, it persists up to 70~K. This persistence, alongside the $M(T)$ behavior, indicates the presence of stable, non-percolating ferromagnetic clusters (macrospins) well above $T_{\mathrm{C}}$, and supports a scenario of progressive fragmentation of the percolating FM network with increasing temperature.

While exhibiting percolative ferromagnetism (Ga,Mn)N retains key features of conventional ferromagnets, including clear resonance behavior and dominant uniaxial magnetic anisotropy with an easy axis perpendicular to the $c$-axis, attributed to Mn$^{3+}$ single-ion anisotropy. No signatures of static Jahn–Teller distortions are observed, implying a transition to a dynamic regime in the studied temperature range.
Atomistic spin simulations qualitatively reproduce the experimental results and offer insights into the nature of Mn–Mn interactions in this dilute system. While the model captures the main trends, further refinements are required to fully account for local disorder and low-field discrepancies.

Overall, our results confirm that FMR serves as a powerful complementary technique to magnetometry in studying percolating ferromagnetic systems. The coexistence of non percolating ferromagnetic clusters  above $T_{\mathrm{C}}$ provides a deeper understanding of the complex interplay between magnetization dynamics, anisotropy, and percolation effects, with potential implications for spintronic applications even well above the Curie temperature.	

	\section{Acknowledgments}
	The work is supported by the National Science Centre (Poland) through project OPUS 2018/31/B/ST3/03438.

\begin{appendices}

	\section{Magnetometry details}
	\label{App:Mgn}
	
	The recorded signals from the (Ga,Mn)N films are weak,  below  10$^{-6}$~emu at remanence, and significantly smaller than the signals from the substrate at high magnetic fields \cite{Gas:2021_JALCOM}.
	The samples are measured both in perpendicular and in-plane orientations, i.~e. with  $\mathbf{H} \parallel \mathbf{c}$ or $\mathbf{H} \perp \mathbf{c}$, respectively.

The magnetothermal properties are examined in weak static magnetic fields using standard field-cooled (FC) and thermo-remanent magnetization (TRM) procedures. 
Special care is taken to ensure an extremely weak residual magnetic field (estimated to be approximately 0.15~Oe) during the TRM measurements.
To achieve this condition, the entire system, including the sample, is degaussed at 300 K prior to the TRM measurements by applying an oscillating magnetic field gradually reduced to zero.
Subsequently, the sample is field cooled from 300 to 2 K.
We underline here that the near-zero field conditions at the base temperature, $T = 2$~K, are established by a soft quench of    the MPMS's superconducting magnet.
This procedure is carried out using the \textsl{magnet reset option} of the MPMS-XL system.	

In this respect (Ga,Mn)N and other strongly dilute compounds in which randonly distributed magnetic species are coupled by a short range interaction constitute a class on their own, for which the correct methodology of $T_{\mathrm{C}}$ determination is yet to be firmly established.
Earlier studies of criticality in (Ga,Mn)N showed that all the traditional magnetometric approaches, such as the temperature dependence of: AC-susceptibility, the inverse of the static susceptibility, disappearance of the coercive field $H_C$, as well as of the thermoremnant magnetization give nearly the same values of the critical temperature \cite{Stefanowicz:2013_PRB,Gas:2018_JALCOM}.
	Some small discrepancies among these values are caused by a smeared character of the Curie transition in dilute systems, either due to the formation of magnetic clusters \cite{Griffiths:1969_PRL} or to the presence of a nonzero variance in the Mn distribution \cite{Stefanowicz:2013_PRB}.

	\section{Simulation details}
		\label{App:Simul}
	
In this Appendix, we provide detailed descriptions of the numerical simulation methodology outlined in the main text, including the complete Hamiltonian, the equations of motion, and the parameter values used in the simulations.
The experimental results are modeled in the frame of an atomistic spin model \cite{Evans:2014_JOP} supplemented with stochastic Landau-Lifshitz-Gilbert (sLLG) equation. 
We use a simulation box of about 25 × 25 × 5 nm$^3$ consisting of approximately 124820 wurtzite lattice cation sites on which the dynamics of 9860 randomly distributed Mn spins ($x=7.9\%$) is simulated. 
In-plane periodic boundary conditions are implemented to model a large thin film. 
We follow the procedure presented in \cite{Sztenkiel:2025_CM}.

We use the following Hamiltonian:
\begin{multline}
\label{Eq:Hamiltonian}
\mathcal{H} = 
\underbrace{- \sum_{\langle i,j \rangle} J_{ij} \pmb{S}_i \cdot \pmb{S}_j}_{\text{Exchange}} 
\overbrace{-\frac{1}{4} K^{TR} \sum_{i} \left[ S_{iz}^2 - \left( S_{ix}^2 + S_{iy}^2 \right) \right]}^{\text{Trigonal}} \\
\underbrace{-\frac{1}{2} K^{JT} \sum_{i} \sum_{j=A,B,C} \left( \pmb{S}_i \cdot \pmb{e}^{JT}_j \right)^4}_{\text{Jahn-Teller}} 
\overbrace{- \mu_S \sum_i \pmb{H} \cdot \pmb{S}_i}^{\text{Zeeman}},
\end{multline}
which consists of  ferromagnetic superexchange interaction between Mn ions, two terms describing the magnetocrystalline energy,  and  the Zeeman energy.
Here, $\mathbf{S}_i$ describes the normalized local magnetic moment of Mn ion located at site $i$ with magnitude $\mu_S=gS\mu_B$, where $g=2$, $S=2$ and $\mu_B$ is the Bohr magneton. 
Mn ions are coupled through a ferromagnetic superexchange interaction of Heisenberg form. 	
We take into account interactions up to the $\mathrm{14^{th}}$ nearest-neighbors approximated by  $J_{ij} = J_0\exp(-R_{ij}/b)$ \cite{Bonanni:2021_HB}, with $R_{ij}$ being the distance between ions $\mathbf{S_i}$ and $\mathbf{S_j}$.

The magnetocrystalline energy is composed of (uniaxial) trigonal and Jahn-Teller anisotropies. 
The latter is approximated by a cubic form  \cite{Edathumkandy:2022_JMMM, Sztenkiel:2025_CM}, with anisotropy axes aligned along the three JT distortion axes, defined after Ref.~\onlinecite{Gosk:2005_PRB}: $e_{A}^{JT}=\left[ \sqrt{\frac{2}{3}}, 0, \sqrt{\frac{1}{3}}  \right]$, $e_{B}^{JT}=\left[-\sqrt{\frac{1}{6}}, -\sqrt{\frac{1}{2}} ,\sqrt{\frac{1}{3}}  \right]$, $e_{C}^{JT} = \left[-\sqrt{\frac{1}{6}}, \sqrt{\frac{1}{2}} ,	\sqrt{\frac{1}{3}}  \right]$.

All the numerical results presented in this study are obtained using the following parameter set: $J_{nn}=4$~meV, $b=1.1$~\AA, $K^{TR}$ = 0.05 meV/atom, $K^{JT}$ = 0.75 meV/atom, and a Gilbert damping constant  $\alpha_G = 0.1$.
 These values were chosen to achieve the best possible agreement with the experimental magnetization curves $M(H)$ for both anisotropy directions, as shown in Fig.~\ref{Fig:REM}~(b).

To calculate the magnetization as a function of external magnetic field $\mathbf{M}(H)$ at a nonzero temperature we initialize the system at a high magnetic field in a completely random spin state. 
The spin system evolves according to the sLLG equation \cite{Edathumkandy:2022_JMMM,Sztenkiel:2025_CM}. 
\begin{equation}
\label{eq:LLG}
\frac{\partial{\textbf{S}_i}}{\partial t}=- \frac{\gamma}{1+\alpha_G^2}
[\textbf{S}_i\times\textbf{H}_{\text{eff}}^{i,tot}+\alpha_G\textbf{S}_i\times(\textbf{S}_i\times\textbf{H}_{\text{eff}}^{i,tot})],
\end{equation}
where $\gamma$ and $\alpha_G$ are the gyromagnetic ratio and the precession damping term, respectively. 
The total effective magnetic field acting on $i$-th spin $\textbf{H}_{\text{eff}}^{i,tot}$ consists of the Zeeman, exchange and anisotropy field $\textbf{H}_{\mathrm{eff}}^i=-(1/\mu_S)\partial\mathcal{H}/\partial\textbf{S}_i$ and the thermal field $\textbf{H}_{th}^i$ (\cite{Skubic:2008_JOP,Evans:2014_JOP,Evans:2015_PRB}. 
Here $\textbf{H}_{th}^i$ represents thermal fluctuations (a nonzero temperature) in the system and is modeled as random magnetic field generated in every sLLG iteration according to:		
\begin{equation}
\textbf{H}_{th}^i= \mathbf{\Gamma} (t) \sqrt{\frac{2 \alpha_G k_B T}{\gamma
		\mu_S \Delta t}},
\end{equation}
where $\mathbf{\Gamma}(t) $ is a 3D Gaussian random vector with a standard deviation of 1 and mean of zero, and $\Delta t = 5 \cdot10^{-6}$~ns is the integration time step. 
The simulation is advanced over time and $\mathbf{M}$ is computed after the system has reached a steady state. 
This steady-state spin configuration then serves as the initial condition for the subsequent simulation step with a slightly reduced $\mathbf{H}$. 
The iterative reduction of $\mathbf{H}$ and corresponding steady-state measurements enable us to construct a detailed $\mathbf{M}(H)$ curve.

To reproduce the TRM data—that is, to calculate the temperature dependence of $M$ at zero magnetic field—we initialize the same numerical box at $T=0.1$~K,  with all Mn spins aligned within the easy plane, i.e., perpendicular to the crystallographic $c$-axis. 
This configuration reflects the physical state of the (Ga,Mn)N film after field cooling to the base temperature of the SQUID magnetometer.
Subsequently, a very weak probing field of $H = 1$~Oe is applied along the direction of the initial spin alignment, and the system—consisting of nearly 10,000 Mn spins—is allowed to relax toward thermal equilibrium.
It is important to note that magnetization-versus-field calculations in the hysteresis region inherently correspond to far-from-equilibrium conditions, involving metastable spin configurations. Therefore, the sLLG formalism must be used to capture the dynamics of spin reversal. 
In contrast, temperature-dependent magnetization data such as $M_{\mathrm{TRM}}(T)$ can be modeled within the framework of thermodynamic equilibrium, which justifies the use of Monte Carlo simulations in this case \cite{Evans:2014_JOP,Sztenkiel:2023_ATS}.
Technically, the system is evolved with a sufficient number of Monte Carlo steps per site (MCS/site) \cite{Zhang:2021_CMS}, where for each MCS, a trial spin orientation change is proposed, and the corresponding energy change $\Delta E$ is calculated. 
The trial move is accepted or rejected based on the Metropolis criterion. 
The system is considered to have reached equilibrium once the magnetization stabilizes and no significant fluctuations are observed. 
After determining the resulting $M$, similarly to the TRM measurement, the simulation proceeds with the temperature increment.
Here, the previous spin configuration is used as the initial condition for the next step.
The procedure is repeated until the full $M_{\mathrm{TRM}}(T)$ curve is established \cite{Prudnikov:2016_JP}.

\begin{figure}
	\centering
	\includegraphics[width=0.48\textwidth]{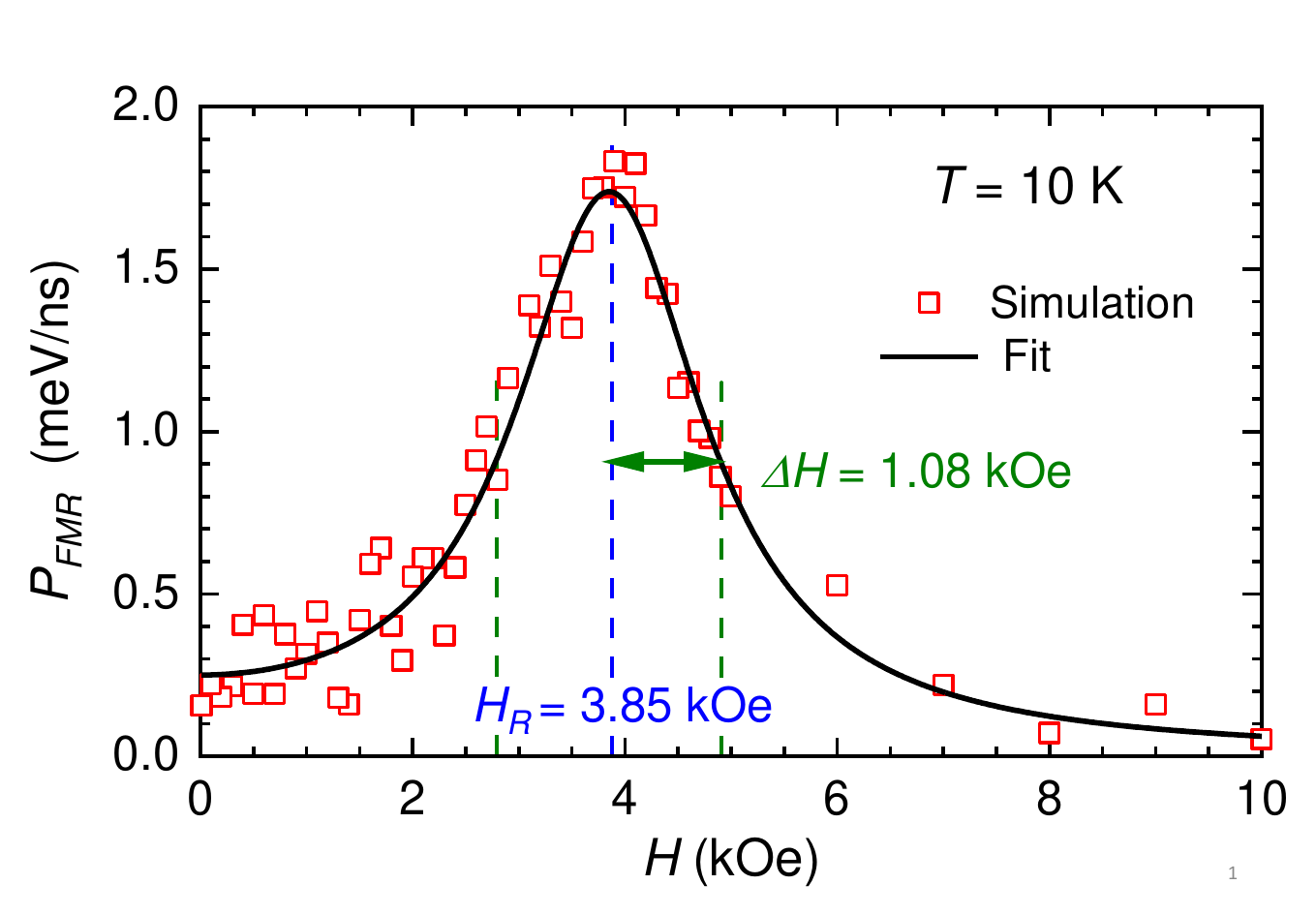}
	\caption{ \label{Fig:SimFRM}
		(Color online) Open squares show an example of the simulated ferromagnetic resonance spectrum, obtained within the same framework described here. The solid line represents a fit using a collision-broadened Lorentzian lineshape function (Eq.~\ref{eq:FitFun}), yielding the resonance field $H_R$ and linewidth $\Delta H$ for this particular simulation.
	}
\end{figure}
To model FMR, we add a sinusoidal $AC$ magnetic field ($\mathbf{h}_{AC}$) with a frequency $\nu = 9.54$~GHz to the external field $\mathbf{H}$ ensuring that $\mathbf{h}_{AC} \perp \mathbf{H}$ \cite{Usadel:2006_PRB}.
First, we start with a completely random spin orientation and a very high magnetic field $H = 60$~kOe. 
The system is allowed to evolve within the sLLG frame until it reaches a steady state, at which we calculate the microwave power absorbed during FMR: $P_{\mathrm{FMR}}=-\frac{\mu_S}{N t_0} \sum_i\int_{t}^{t + t_0} \mathbf{S_i} \frac{\partial \mathbf{h}_{AC}(t)}{\partial t} dt$, where $N$ is the number of Mn ions and $t_0$ is simulation time \cite{Usadel:2006_PRB}. 
Next, this steady-state spin configuration is used as the initial condition for the next simulation step, in which the external field $\mathbf{H}$ is slightly reduced. 
This process is repeated iteratively,  until $\mathbf{H}$ reaches zero. 
Exemplary $P_{\mathrm{FMR}}(H)$ simulation data are depicted in Fig.~\ref{Fig:SimFRM} by open symbols.

The resonance field $H_R$ and the line-width $\Delta H$ of the simulated FMR are determined at the given $T$ by fitting the collision-broadened Lorentzian line-shape function $f(H)$ \cite{Vanvleck:1945_RMP}:
\begin{equation}
\label{eq:FitFun}
f(H)=\frac{A}{\pi} \left[ \frac{\Delta H}{(H - H_R)^2 + \Delta H^2}
+ \frac{\Delta H}{(H + H_R)^2 + \Delta H^2} \right]
\end{equation} 
to the modeled $P_{\mathrm{FMR}}(H)$. 
A result of such a fit is represented by solid line in Fig.~\ref{Fig:SimFRM}.
The entire process is repeated for temperatures ranging from 2 to 22~K.

Technically, in the sLLG simulations, the code is parallelized by dividing the simulation region into sections, with each processor handling a specific part of the system. 
Computations are accelerated using graphics processing units (GPUs), and Euler’s method is employed as the integration scheme after checking that it yields the
same results as the Heun’s scheme (see Appendix~\ref{App:Integr}), and that all results are stable.
Since a higher temperature introduces more disturbance in the system, the initialization and averaging steps of the sLLG simulations are adjusted as specified in table~\ref{tab:mono_mn_abc}.
\begin{table}[htb]
	\caption{\label{tab:mono_mn_abc} Parameters of atomistic spin model simulations. }
	\begin{ruledtabular}
		\begin{tabular}{ccc}
			Temperature	  & Number of  & Number of \\
			(K)   &		initialization steps		 & 	averaging steps		 \\
			\colrule \\
			0.01 to 0.40& $1\times10^7$ & $1\times10^7$ \\
			0.60 		& $1\times10^7$ & $3\times10^7$ \\
			0.80 to 1.00& $1\times10^7$ & $4\times10^7$ \\
			1.50		& $1\times10^7$ & $6\times10^7$ \\
			2.00		& $2\times10^7$ & $5\times10^7$ \\
			2.50		& $2\times10^7$ & $6\times10^7$ \\
			3.00 to 3.50& $2\times10^7$ & $8\times10^7$ \\
			4 to 22		& $6\times10^7$ & $6\times10^8$ \\
		\end{tabular}
	\end{ruledtabular}
\end{table}

\section{Integration scheme validation}
\label{App:Integr}

\begin{figure*}[htb]
	\centering
	\includegraphics[width=0.98\textwidth]{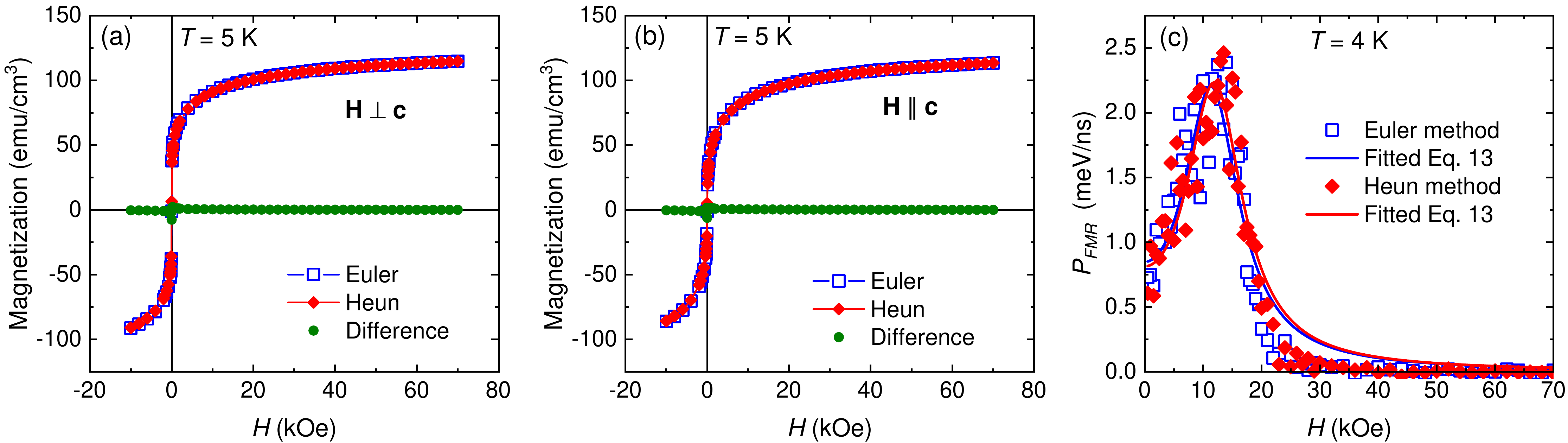}
	\caption{(Color online) Comparison between Euler (open rectangles) and Heun (full circles) integration schemes used during stochastic Landau–Lifshitz–Gilbert simulations for: (a) magnetization curves $M(H)$ at $T=5$~K for $\mathbf{H} \perp \mathbf{c}$, (b) $M(H)$ for $\mathbf{H} \parallel \mathbf{c}$, and (c) FMR spectrum at $T=4$~K with $\mathbf{H} \parallel \mathbf{c}$. The green diamonds in panels (a) and (b) mark the difference between the results of the two integration schemes, while the solid lines in panel (c) shows the results of the fit of the Eq.~\ref{eq:FitFun} to the numerically simulated points.}
	\label{Fig:EulerHeun}
\end{figure*}

In the sLLG simulations the Euler integration scheme has been applied.
To ensure that our choice  does not affect the reliability of the modeling results, we perform a comparative analysis with the Heun method. 
Simulations are conducted using identical parameters and timestep $\Delta t = 5 \cdot 10^{-6}$~ns for both schemes. 
Fig.~\ref{Fig:EulerHeun} shows a side-by-side comparison of magnetization curves and FMR spectra obtained by both methods.

As illustrated, both the static magnetization and dynamic FMR results are in very good agreement for the two schemes. 
The relative differences in resonance field ($H_R$) and linewidth ($\Delta H$) remain within 5\%.

\section{Limitation of the model}
\label{App:Limitations}

While the atomistic spin model employed in this study captures the essential features of the magnetization behavior in (Ga,Mn)N, several simplifying assumptions have been applied, which limit the accuracy of the simulations—particularly in the low-field regime where the Zeeman energy is insufficient to overcome subtle anisotropy effects. 
The primary limitations of the current model are as follows:
\begin{itemize}
	\item 
Neglect of Mn–Mn pair anisotropy:
Pairwise anisotropic interactions between neighboring Mn ions are not included in the present model. These interactions have been shown to influence magnetic behavior in dilute systems and are currently under intensive investigation in a follow-up study \cite{Das_PairAnizo_arXiv}.
These effects may be important, as, assuming purely statistical Mn distribution, for $x=0.08$ as much as  $1-(1-x)^{12}$ Mn ions, that is 63\% of all ions must have at least one nearest-neighbor.  
Their inclusion is expected to improve the fidelity of the simulated remnant magnetization and coercivity.
	\item
Cubic approximation of Jahn–Teller anisotropy:
In reality, the Jahn–Teller effect leads to a tetragonal distortion along one of the cubic ⟨100⟩ axes, with the trigonal ⟨111⟩ axis aligned with the crystallographic $\mathbf{c}$-axis. 
This implies that the distortion is active along \textsl{only} one of the local axes  at a time ($e_{A}$, $e_{B}$, or $e_{C}$ - defined in the previous section), as captured in direct quantum mechanical calculations and crystal field model \cite{Wolos:2004_PRB_69, Gosk:2005_PRB}. 
However, such directional specificity (lowering of the site symmetry) cannot be implemented within the classical sLLG framework. 
Therefore, we approximate the Jahn–Teller contribution using an effective cubic anisotropy term (the second part of Eq.~\ref{Eq:Hamiltonian}), i.e., acting equally along the three crystallographic directions, in contrast to the Jahn–Teller distortion, which selects only one axis at a time.
	\item
Assumption of a perfectly random Mn distribution:
The spatial distribution of Mn ions is assumed to be statistically random, without accounting for clustering or local concentration fluctuations that may occur during growth \cite{Gonzales:2011_PRB}. 
Experimental and theoretical studies suggest that such inhomogeneities can significantly modify the local anisotropy landscape and spin connectivity, particularly in the dilute regime \cite{Stefanowicz:2013_PRB, Gas:2018_JALCOM}.
\end{itemize}
Among the above effects, only the pairwise anisotropy is currently being implemented, with preliminary results forming the basis of a separate ongoing study \cite{Das_PairAnizo_arXiv}. 
The inclusion of the remaining contributions—namely, the directional nature of the Jahn–Teller distortion and deviations from random Mn distribution—requires more substantial model development, and is planned for future work.

\end{appendices}

\bibliography{references.bib}

\end{document}